\documentclass[twocolumn,showpacs,amsmath,amstex,amssymb,mathfonts,prl]{revtex4-1}
\usepackage{graphicx}

\usepackage{tipa}
\usepackage{bm}

\usepackage{amsmath}
\usepackage{amssymb}
\usepackage{amsthm}
\usepackage{amsfonts}
\usepackage{float}

\begin{document}

\title{Classification and Unification of the Microscopic Deterministic Traffic Models}

\author{Bo Yang and Christopher Monterola}
\affiliation{Complex Systems Group, Institute of High Performance Computing, A*STAR, Singapore, 138632.}
\date{\today}

\date{\today}
\begin{abstract}
We identify a universal mathematical structure in microscopic deterministic traffic models (with identical drivers), and thus show that all such existing models in the literature, including both the two-phase and three-phase models, can be understood as special cases of a master model by expansion around a set of well-defined ground states. This allows any two traffic models to be properly compared and identified. The three-phase models are characterised by the vanishing of leading orders of expansion within a certain density range, and as an example the popular intelligent driver models (IDM) is shown to be equivalent to a generalized optimal velocity (OV) model. We also explore the diverse solutions of the generalized OV model that can be important both for understanding human driving behaviours and algorithms for autonomous driverless vehicles. 
\end{abstract}

\maketitle 

It is of great interest, both theoretically and practically, to understand via simple models the emergent behaviours of the complex systems containing a large number of interacting components. Examples like crowd dynamics\cite{helbing_crowd,helbing_nature}, highway traffic systems\cite{dogbe, as}, and the more recent urban traffic flows\cite{helbing_control} have attracted physicists for decades. Despite the observed complex patterns from these systems, some of the essential characteristics are universally well-defined\cite{helbing, kernerbook}. In contrast to the traditional many-body physical systems (e.g.  involving identical particles like electrons), the crowd or traffic systems lack almost any symmetry at the microscopic level: even individual components are different from one another, with intrinsic stochasticity. This poses great challenges in finding the simplest model with enough predictive power to adequately characterise these complex systems.

The lack of symmetries at the microscopic level leads to a certain arbitrariness in the model construction. The modelling of the highway traffic system has led to a plethora of traffic models\cite{dogbe, as, helbing, kernerbook}. Analysis of the empirical data has led to profound understanding of the traffic dynamics\cite{kernerexp1,kernerexp2}; however due to non-linear interactions between drivers, simple interactions can lead to complex spatiotemporal patterns. It is thus difficult to decide which detailed driving behaviours or specific functional forms should be used in the model. This is the primary concern behind the controversy between the two-phase and three-phase traffic theories\cite{kernercrit,helbingcrit}, which leads to two classes of traffic models. While both the two-phase and the three-phase traffic models can simulate the scattering of the data points on the flow-density plane when the traffic is congested, the two classes are distinguished by the presence or absence of a continuous fundamental diagram in the flow-density plot\cite{kernerbook}. Moreover, the lack of a fundamental principle in understanding various different traffic models makes it difficult to decide if, in addition to the non-linear interactions between the vehicles, factors like stochasticity or diversity of driving behaviours are also essential to certain observed empirical features.

To tackle the dilemma of the apparent (over)abundance of traffic models, we need answers to the following questions: a). How do we properly characterise the differences between two traffic models? b). Is there a standard way of extending an existing traffic model or construction of new traffic models? c). Is there a standard way in selecting the best traffic model based on the empirical data? In this Letter, we answer in details the first two questions for deterministic microscopic models with identical drivers, by proposing a general framework of obtaining all such models from a master model. The controlled expansion around properly defined ``ground states" of the master model shows that the two-phase and three-phase traffic models are both just special cases. The general framework also allows us to classify existing and new traffic model, and compare explicitly various types of approximations in a well-defined way.

The answer to the third question lies in the fact that in principle the master model can be obtained empirically, via a renormalization-like procedure by averaging over unimportant factors influencing the reaction of individual drivers. The details can be found in\cite{yangbo} and will also be discussed elsewhere. We will now develop the general framework by assuming the following form of the master model
\begin{eqnarray}\label{master}
\tilde a_n=\tilde f\left(\tilde h_n,\tilde v_n,\Delta \tilde v_n\right)
\end{eqnarray} 
Here the assumption is that the bumper-to-bumper distance of the $n^{\text{th}}$ vehicle, $\tilde h_n$, its velocity $\tilde v_n$, and the approach velocity $\Delta \tilde v_n=\tilde v_{n+1}-\tilde v_n$ are the important factors affecting the acceleration of the $n^{\text{th}}$ vehicle, and all other factors are averaged over. One can choose more (or different) dynamic variables in the function $\tilde f$, but for the purpose of illustration we choose this simple yet sufficient case. 

The natural time and length scale of the traffic system is given by the two statistically robust empirical quantities: $\rho_j$, the density of the vehicles within a wide moving jam, and $V_{\text{max}}$, the maximum velocity (or the speed limit). Since we will carry out a controlled expansion of Eq.(\ref{master}) later, all quantities in Eq.(\ref{master}) are dimensionless after a chosen set of scales. While formally the model and its expansion are independent of the choice of the scales, for practical purposes one can choose either the length scale $\rho_j^{-1}$, the time scale $\left(\rho_jV_{\text{max}}\right)^{-1}$, or other sets of the physically relevant scales, depending on different purposes and the particular system under study.

\begin{figure}
\begin{center}
\includegraphics[height=6cm]{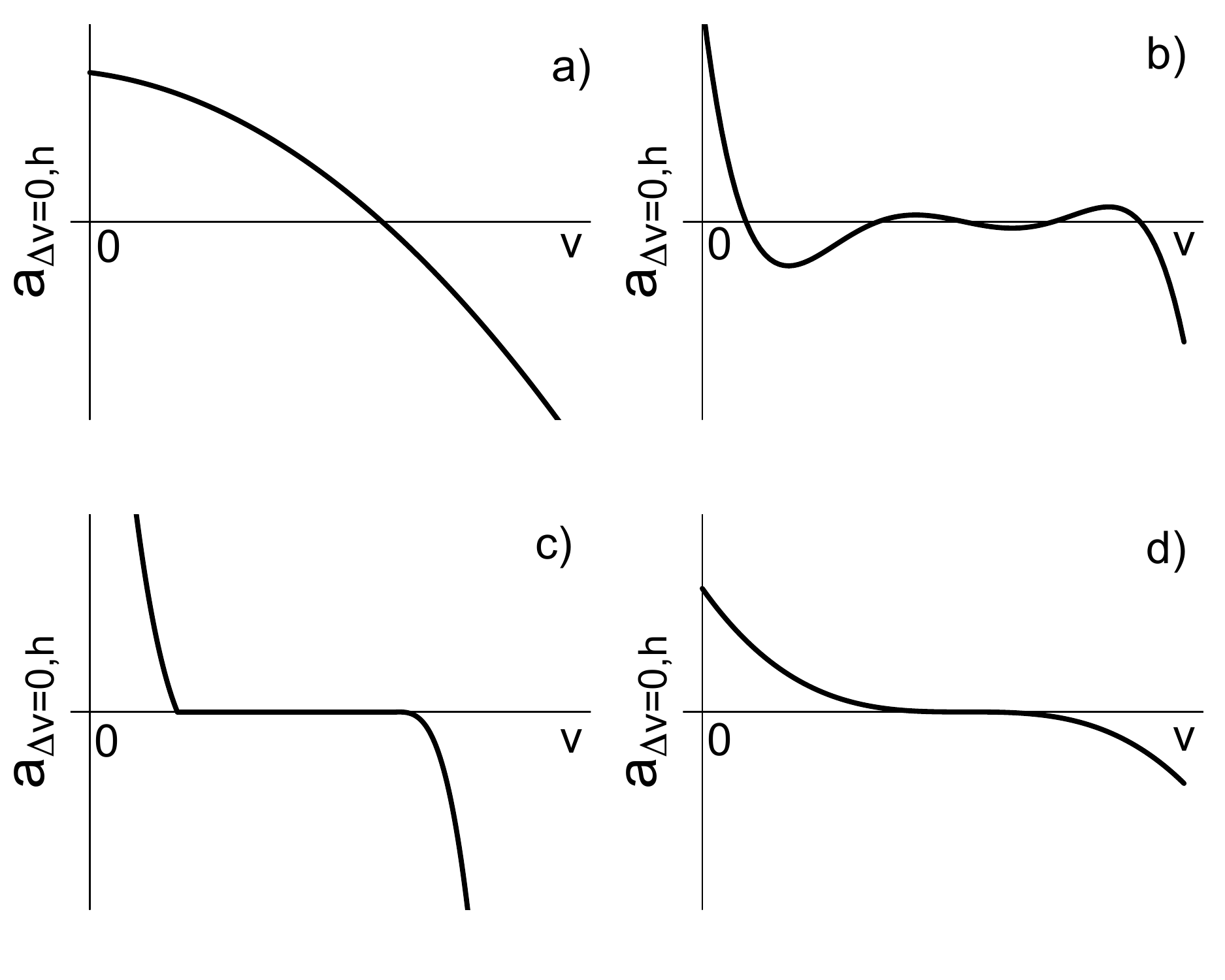}
\caption{Various possible sections of the acceleration (given by $\tilde f$ in Eq.(\ref{master})) at different $h$, as a function of $v$ and evaluated at $\Delta v=0$. The non-analyticity of c) can be well approximated by d). The specific form of such acceleration can in principle be obtained from empirical data\cite{yangbo}}.
\label{fig1}
\end{center}
\end{figure}

The properties of the master model can be studied by plotting $\tilde f\left(h,v,\Delta v\right)$ as a function of $v$ at fixed $h$ in Fig.(\ref{fig1}), where we take $\Delta v=0$ and focus around zero accelerations. The master model has a fundamental diagram if for any value of $h$ the acceleration decreases monotonically with $v$ and crosses the x-axis only once, as shown in Fig.(\ref{fig1}a) or Fig.(\ref{fig1}d). This corresponds to the two-phase traffic models with a unique relationship between the flow and density. The difference between two-phase models as reflected by the detailed plots can be quantified by the proper Taylor expansions as we will show later. 

The plots in Fig.(\ref{fig1}b) and Fig.(\ref{fig1}c) are cases where for certain range of $\tilde h$, there are more than one steady state (given by the vanishing acceleration) at the same density ($\sim\tilde h^{-1}$), but with different average velocities (given by the x-axis intersections). They correspond to the microscopic three-phase models proposed in the literature\cite{kerner_mod1,kerner_mod2}. For Fig.(\ref{fig1}b), the number of steady states at the same density equal to the number of intersections at the x-axis, though only the ones with negative gradient are stable. The Taylor expansions around different stable points where $\tilde a=0$ leads to different optimal velocity functions within different velocity range. If for certain range of $\tilde h$ the acceleration is not analytic in $\tilde v$ as shown in Fig.(\ref{fig1}c), steady states can occur at the same average traffic density but for any velocities at which the acceleration is zero. Since the plot corresponds to the case where $\Delta \tilde v=0$, this is exactly the case of speed adaptation mechanism that leads to the scattering of the flow-density plot at the synchronized phase. Both these two cases are commonly used in various forms for the construction of three-phase models.

The key assumption of the three-phase traffic theory is that the scattering of the flow-density plot in the ``synchronized phase" corresponds to a multitude of \emph{steady} states with non-unique flow-density relations\cite{kernerbook}. it is however empirically difficult to verify if those states are indeed steady states. The macroscopic quantities like flow and density are empirically obtained from the aggregated raw data\cite{treiberbook,helbingcrit}. A transient state with qualitative or quantitative features that evolves very slowly will be captured by the sensor as the ``steady state". Thus in most cases we cannot distinguish such long lasting quasi-steady states from the steady states. 

We thus argue that instead of dealing with a non-analytic function as shown in Fig.(\ref{fig1}c), we can for most practical purposes replace it with an analytic function with an inflection point at zero acceleration (i.e. Fig.(\ref{fig1}d)). Indeed the function in Fig.(\ref{fig1}c) can be approximated by an analytical function to any degree of accuracy. Such a model still has a fundamental diagram, but for any state not too far away from the inflection point, it evolves very slowly with very small accelerations. Numerical simulation of such analytic models show very similar scattering of the flow-density plot and the spatiotemporal patterns as the proposed three-phase models within any physically reasonable time scale. Therefore fundamentally there is no strict boundary between the two-phase and three-phase models, and we will proceed to show the way to differentiate them is \emph{no different} from the way of differentiating different two-phase models.

We will now only deal with analytic master models. The ground states of the model are defined as the solutions with all vehicles equally spaced apart with headway $\tilde h$ and travelling at the same time-independent velocity $\tilde v$. In Eq.(\ref{master}) the distance (or headway) is scaled by $\rho_j^{-1}$ while the time is scaled by $\rho_j^{-1}V_{\text{max}}^{-1}$; the tilded velocities and accelerations are derivatives of the scaled distance by the scaled time. Thus all quantities with tilde are scaled to be dimensionless. The ground states are indexed by the average density $\tilde h^{-1}$, and at some densities there may be more than one ground states (i.e. Fig.(\ref{fig1}b)). We can thus expand the master model around a chosen ground state:
\begin{eqnarray}\label{texpand}
&&\tilde a_n=\sum_{p,q}\kappa_{p,q}\left(\tilde h_n\right)\left(\tilde v_n-V_{op}\left(\tilde h_n\right)\right)^p\Delta\tilde v_n^q\\
&&\kappa_{p,q}\left(\tilde h_n\right)=\frac{1}{p!q!}\frac{\partial^{p+q}\tilde f}{\partial^p\tilde v_n\partial^q\Delta\tilde v_n}\bigg|_{\begin{subarray}{l}\tilde v_n=\tilde V_{op}\left(\tilde h_n\right)\\\Delta\tilde v_n=0\end{subarray}}
\end{eqnarray}
where $V_{op}\left(h\right)$ satisfies $\tilde f\left(h,V_{op}\left(h\right),0\right)=0$. Clearly any existing two-phase models can be expanded this way, and they can be quantitatively compared by the set of coefficients $\kappa_{p,q}$. The optimal velocity (OV) model\cite{bando} is the special case where $\kappa_{1,0}$ is a negative constant while all other $\kappa_{p,q}$ vanishes. The related FVD model has an additional constant $\kappa_{0,1}$\cite{JiangR_PRE01}, while the asymmetric FVD\cite{GLW_PhyA08} can be understood as having additional non-linear corrections with non-vanishing $\kappa_{0,q}, q>1$. 

Within this framework the three-phase models and the phase transition can be understood as follows: there is a certain range of the density (typically when the bumper-to-bumper distance is smaller than some ``synchronization gap" but larger than that within a wide moving jam), $\kappa_{p,0}$ vanishes for $p<p_0$ so that the leading terms involving $\tilde v_n$ are of higher orders. The larger the value of $p_0$, the more long lasting the transient states are around the inflection points. Thus if one can obtain $\tilde f$ empirically, the assumptions of the three-phase traffic theory can be tested microscopically.

\begin{figure}
\begin{center}
\includegraphics[height=7cm]{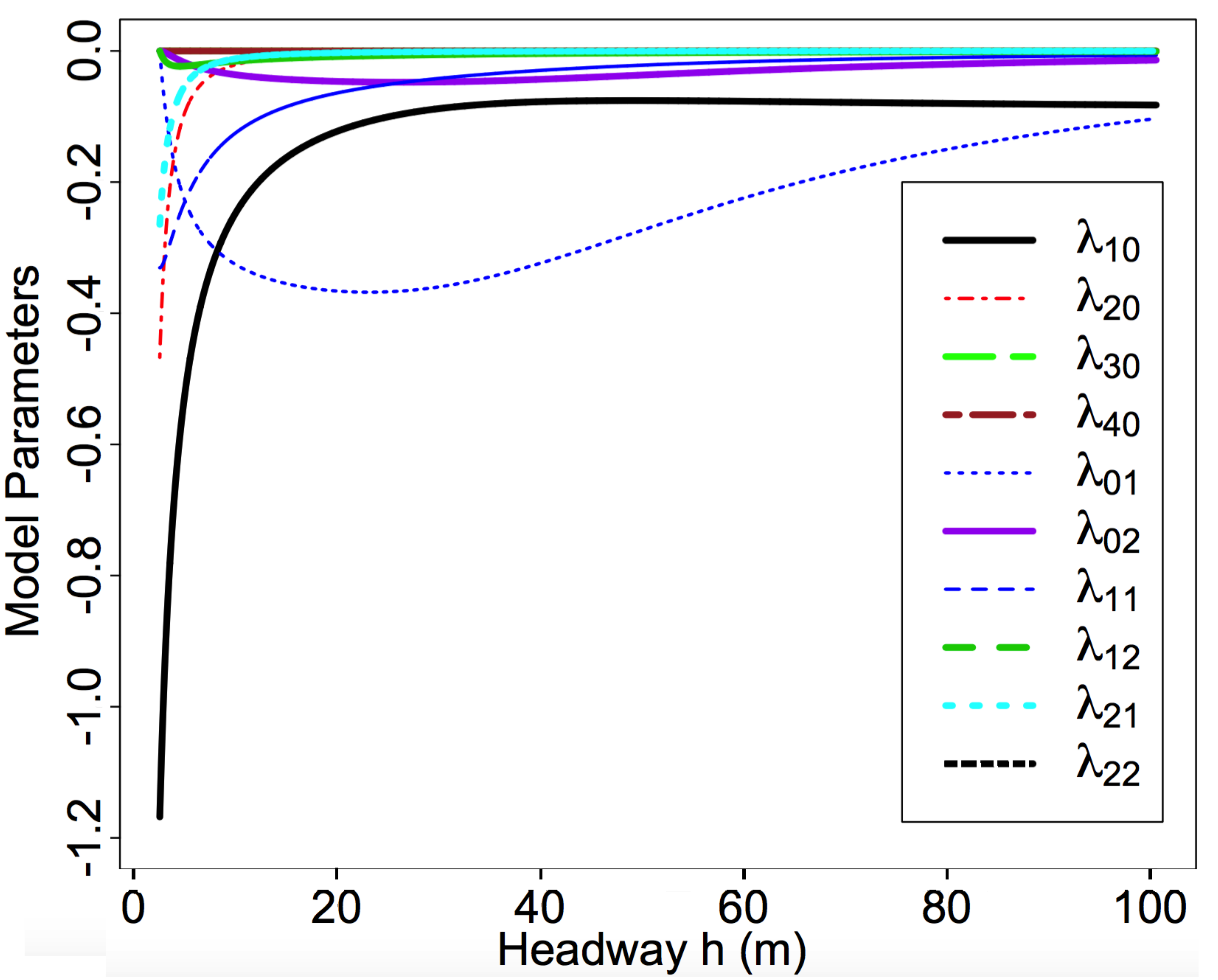}
\caption{(Color Online.)The coefficients of expansion of the IDM  in Eq.(\ref{idm}) as the function of the bumper-to-bumper distance $h$, and $\lambda_{pq}$ is defined in Eq.(\ref{idmtaylor}) and normalised to have the same dimensions}.
\label{fig2}
\end{center}
\end{figure}

Seemingly different traffic models can be shown to be qualitatively equivalent under this scheme of controlled expansion. As an example we look at the IDM model which has the following form:
\begin{eqnarray}\label{idm}
&&a_n=a\left(1-\left(\frac{v_n}{v_0}\right)^\delta-\left(\frac{h^*\left(v_n,\Delta v_n\right)}{h_n}\right)^2\right)\\
&&h^*\left(v,\Delta v\right)=s_0+s_1\sqrt{\frac{v}{v_0}}+Tv+\frac{v\Delta v}{2\sqrt{ab}}
\end{eqnarray}
Here we use dimensionful and physical quantities, with the standard set of parameters\cite{helbing_idm} $a=0.73 ms^{-2}, b=1.67 ms^{-2}, v_0=33ms^{-1}, s_0=2m, s_1=0m, T=1.6s, \delta=4$. The ground state is unique for any density; the relationship between $h$ and $V_{op}$ is given by
\begin{eqnarray}\label{idmov}
h=\left(s_0+V_{op}T\right)\left(1-\left(\frac{V_{op}}{v_0}\right)^\delta\right)^{-\frac{1}{2}}
\end{eqnarray}
The Taylor expansion around the ground state configuration has a finite number of terms:
\begin{eqnarray}\label{idmtaylor}
a_n=\sum_{p=0,q=0}^{p=4,q=2}\lambda_{p,q}\left(v_n-V_{op}\left(h_n\right)\right)^p\Delta v_n^q
\end{eqnarray} 
In general the majority of the approach velocities of the highway traffic, as well as the deviation of the vehicle velocities from the optimal velocities, is on the order of $1ms^{-1}$, as verified from the numerical simulation of Eq.(\ref{idm}). We can thus de-dimensionalize velocities in Eq.(\ref{idmtaylor}) in this unit, normalizing all the non-vanishing coefficients of expansion to have the unit of the physical acceleration, and plot them in Fig.(\ref{fig2}) as functions of the density. One can clearly see that the dependence of $\lambda_{pq}$ on traffic density is quite intuitive. The driver is only sensitive to $\Delta v_n$ at intermediate traffic densities. In the congested traffic the drivers are more alert (large $\lambda_{10}$), and to keep a safe bumper-to-bumper distance is more important than to synchronize the velocity.

\begin{figure}
\begin{center}
\includegraphics[height=6cm]{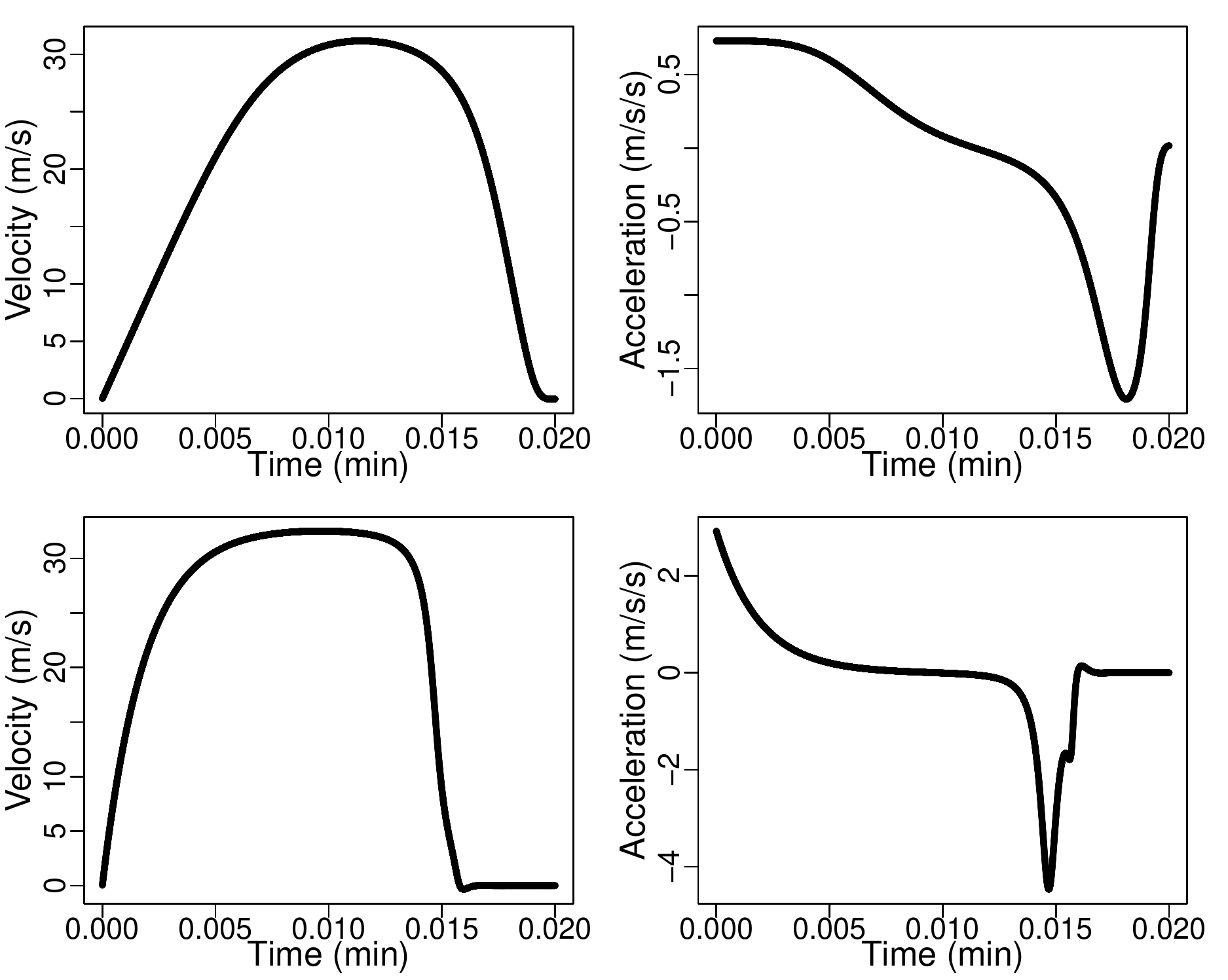}
\caption{Here we simulate the dynamics of a single vehicle accelerating from zero velocity and approaching an obstacle located at $2.5 km$ away from the starting point. The top two plots are the time evolution of the velocity and acceleration, simulated from the full IDM model. The bottom two plots are the time evolution of the velocity and acceleration, simulated from the linearlized IDM model by keeping only $\lambda_{10}$ and $\lambda_{01}$.}.
\label{sup}
\end{center}
\end{figure}

It is thus probably enough from Fig.(\ref{fig2}) to just keep $\lambda_{10},\lambda_{01},\lambda_{11}$ for the qualitative features of the traffic dynamics to be preserved. Numerical simulations show even ignoring $\lambda_{11}$ may be good enough. Such a truncated IDM model can also qualitatively simulate the approach dynamics of a single vehicle rather well (see Fig.\ref{sup}), even though in this special case the truncated terms are presumably not small due to the large approach velocity and large deviation of the vehicle's velocity from its optimal velocity. For the resulting truncated model, further tuning and optimization of the model are possible by fitting $\lambda_{10}$ and $\lambda_{01}$ using simple functions (such as the step-like functions), to improve the quantitative agreements with the original IDM. 

Truncation of higher orders and simplification of the expansion coefficients can be applied to other models like Shamoto's\cite{shamoto}. We would also like to emphasize that it is not \textit{a priori} true that the IDM or Shamato's models are more realistic than their simplified counterparts. The tuning of the expansion coefficients should be based on the experimental measurement of the master model, and it is possible to achieve better agreement with the empirical observations even with the truncated models.

The original OV models\cite{bando, JiangR_PRE01, GLW_PhyA08} are the simplest case of Eq.(\ref{texpand}) where the coefficients of expansion are constants and only the term linear in $\left(v_n-V_{op}\left(h_n\right)\right)$ is kept. The phase diagram of such models are well-known\cite{kerner_phase,yu_phase,naka_phase,nagatani_phase,xue_phase}. To understand the more realistic traffic models as shown above, and for tuning and construction of new traffic models, it is also important to understand the emergent characteristics of the generalized OV models with general coefficients of expansion and higher orders of expansion. This is in general difficult due to the non-linear interaction, and we will just explore some simple examples, leaving a more detailed discussion elsewhere.

 
\begin{figure}
\begin{center}
\includegraphics[height=4.3cm]{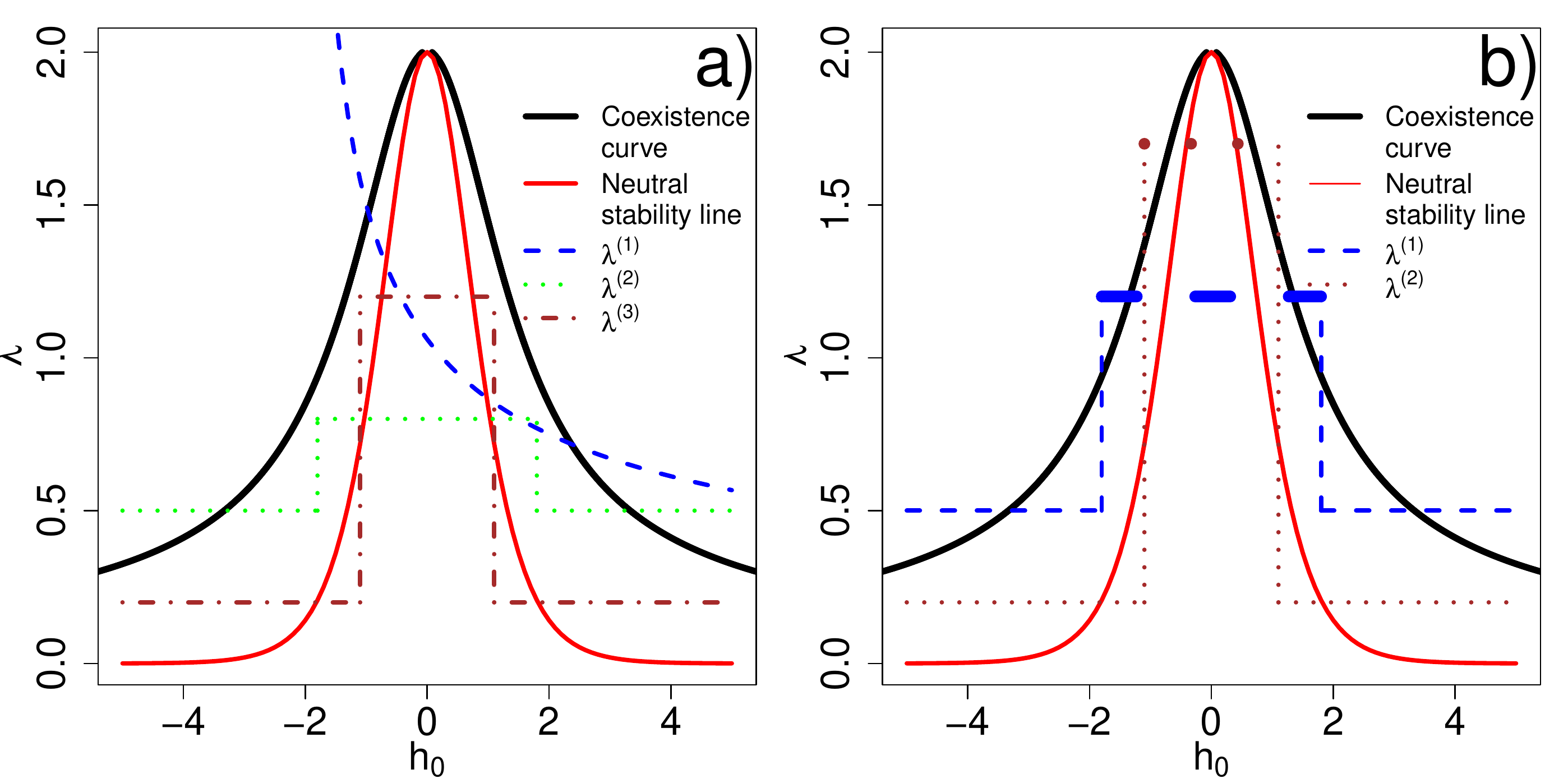}
\includegraphics[height=2.8cm]{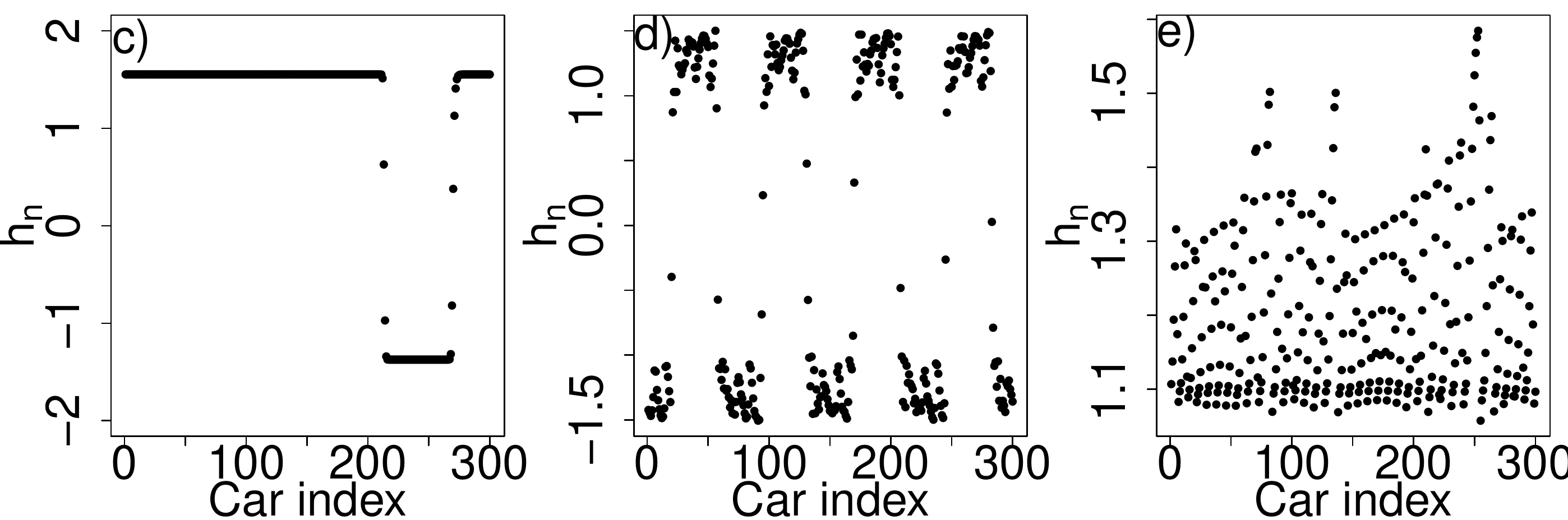}
\caption{(Color Online.)All the insets are plots of headways as a function of the car index. a). $\lambda^{(1)},\lambda^{(2)}$ and $\lambda^{(3)}$ are examples of different functions of $\lambda$ in Eq.(\ref{gov1}). For cluster solutions the maximum or minimum space gaps can be uniform as c)., from $\lambda^{(1)}$ and $\lambda^{(2)}$), or fluctuating as d)., from $\lambda^{(3)}$). b). $\lambda^{(1)}$ and $\lambda^{(2)}$ are examples of different functions of $\lambda$. The vehicle density corresponding to the thickened part of $\lambda$, if unstable, leads to regular cluster solutions. e). shows incomplete cluster solutions when the vehicle density satisfies $\lambda\left(h_0\right)<2\text{sech}h_0$, which is linearly unstable. }.
\label{fig3}
\end{center}
\end{figure}
We start by generalizing the simplest artificial OV model as follows:
\begin{eqnarray}\label{gov1}
a_n=\lambda\left(h_n\right)\left(\tanh\left(h_n\right)-v_n\right)
\end{eqnarray}
It is useful to plot $\lambda$ together with the coexistence curve (CC) as well as the neutral stability (NS) line of the original OV model. The CC at $h<0$ is given by the minimum bumper-to-bumper distance of the cluster solution, which we call the $h_{\text{min}}$-branch; naturally the other half is called the $h_{\text{max}}$-branch\cite{fn}. The linear stability condition at average density $h_0^{-1}$ is given by $\lambda\left(h_0\right)>2\text{sech}^2h_0$. 

The qualitative features of Eq.(\ref{gov1}) can be roughly classified by the way $\lambda$ intersects with the CC and the NS line. If $\lambda$ intersect with the $h_{\text{min}}$- and the $h_{\text{max}}$-branch at most once separately (see Fig.(\ref{fig3}a)), the constant density ($h_n=\text{const}$) solution of Eq.(\ref{gov1}) can be stable or unstable against the development of cluster solutions with minimum and maximum bumper-to-bumper distances $\bar h_{\text{min}}$ and $\bar h_{\text{max}}$, qualitatively the same as the original OV model; both $\bar h_{\text{min}}$ and $\bar h_{\text{max}}$ are \emph{independent} of the density of the traffic. Clearly only $\lambda\in \left[\lambda\left(\bar h_{\text{min}}\right), \lambda\left(\bar h_{\text{max}}\right)\right]$ contributes to the cluster structure, with the following relationship:
\footnotesize
\begin{eqnarray}\label{range}
&&\bar h_{\text{max}/\text{min}}\in\left[h_{\text{max}/\text{min}}\left(\lambda\left(\bar h_{\text{min}}\right)\right),h_{\text{max}/\text{min}}\left(\lambda\left(\bar h_{\text{max}}\right)\right)\right]
\end{eqnarray}
\normalsize
If $\lambda\left(\bar h_{\text{min}}\right)<2\text{sech}(\bar h_{\text{min}})$ or $\lambda\left(\bar h_{\text{max}}\right)<2\text{sech}(\bar h_{\text{max}})$, the vehicles in the clusters (or anti-clusters) are also linearly unstable and their bumper-to-bumper distances will fluctuate (see $\lambda^{(3)}$ in Fig.(\ref{fig3}a) and the inset), in contrast to the original OV model.

If $\lambda$ intersects with the $h_{\text{min}}$-branch ($h_{\text{max}}$-branch) more than once (Fig.(\ref{fig3}b)), the traffic dynamics can be \emph{different} at different vehicle density. For $h_0$ within the thickened part of $\lambda^{(1)}$ and $\lambda^{(2)}$, the stability of the $h_n=\text{const}$ solution and the development of the cluster solution are similar as the cases in Fig.(\ref{fig3}a). For other part of $h_0$, if $\lambda\left(h_0\right)<2\text{sech}(h_0)$, the $h_n=\text{const}$ solution is linearly unstable but clearly the cluster solution cannot be fully developed due to the conservation of the vehicle density. Partially developed cluster solutions with strong oscillations are obtained as shown in the inset of Fig.(\ref{fig3}b)\cite{fn2}. 

\begin{figure}
\begin{center}
\includegraphics[height=7cm]{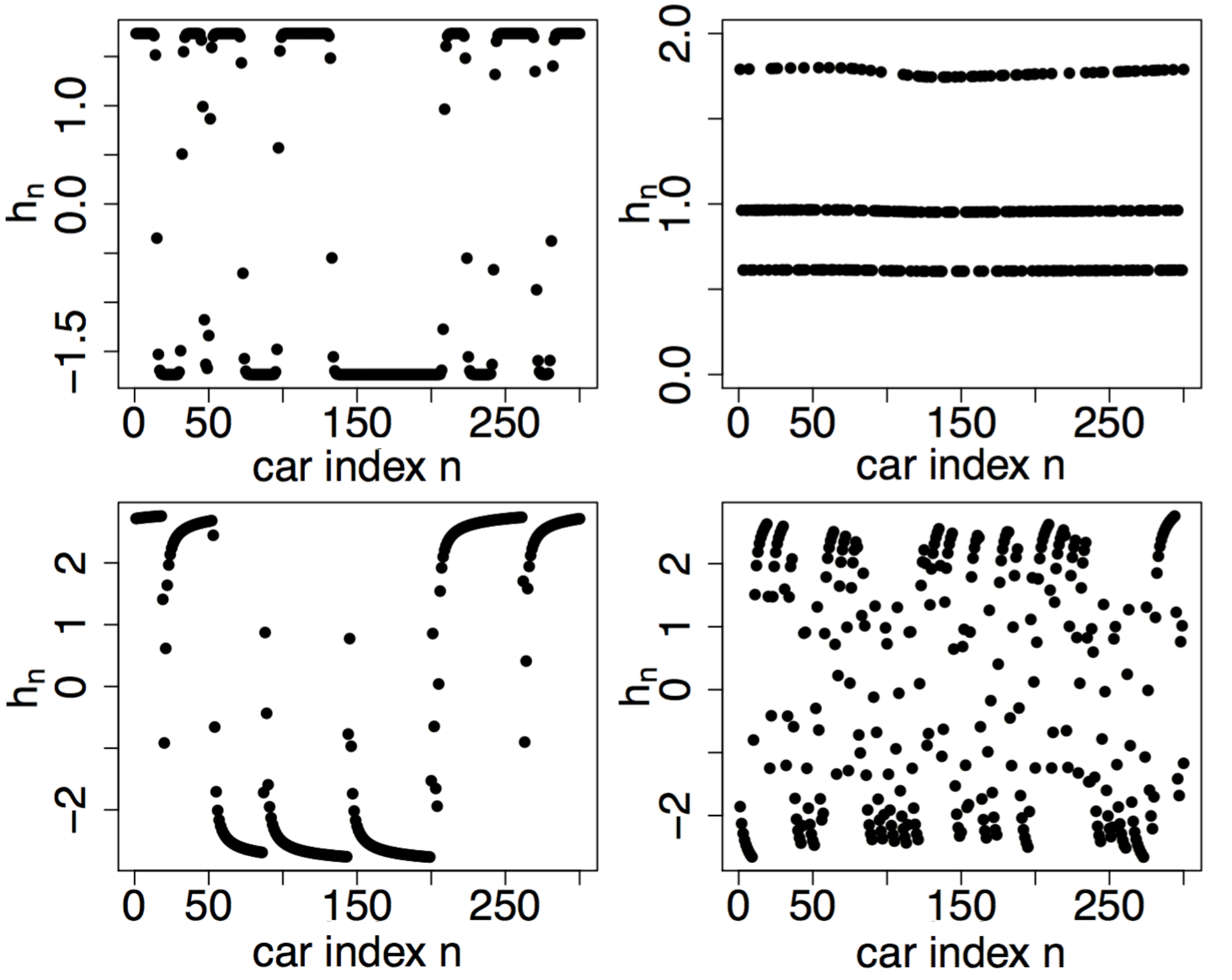}
\caption{a). Cluster solutions in the unstable phase, when higher order terms in $\left(v-V_{op}\right)$ are just small perturbations (Top left). b). Multiple equilibrium densities at the same velocity, corresponding to the case in Fig.(\ref{fig1}b), when the $h_n=\text{const}$ solution at each x-axis intersection with negative gradient is stable (Top right). c). The unstable phase when only $\left(v-V_{op}\right)^3$ term is present in the expansion with constant expansion coefficient (Bottom left). d). The unstable phase when only $\left(v-V_{op}\right)^5$ term is present in the expansion with constant expansion coefficient (Bottom right).}
\label{fig4}
\end{center}
\end{figure}

Higher orders of $\left(V_{op}-v_n\right)$ generally shift the values of $\bar h_{\text{min}}$ and $\bar h_{\text{max}}$. Here we look at simple cases with constant coefficients of expansion. For situations like the ones in Fig.(\ref{fig1}b), the stable phase can have multiple equilibrium densities with vehicles all traveling at the same velocity (see Fig.(\ref{fig4}b)). For three-phase models with vanishing low order terms (at least the linear oder), the NS line is no longer applicable. Since the accelerations are very small even when the vehicles deviates from their equilibrium density, it is very difficult for the clusters to form, as shown in Fig.(\ref{fig4}c) and Fig.(\ref{fig4}d).

One should note that our method can be applied to more general cases, in particular with more realistic optimal velocity functions. The cluster solutions, the CC and NS line are universal features for the more complicated models. Though step functions of $\lambda$ are used in many of the arguments above for its simplicity, $\lambda$ can be deformed moderately or smoothened without altering the qualitative features. Since only $\lambda\in\left[\lambda\left(\bar h_{\text{min}}\right),\lambda\left(\bar h_{\text{max}}\right)\right]$ determines the instability of the model, one can alter the shape of $\lambda$ outside of that range arbitrarily without changing the cluster structure or the phase diagram. For example, the inset of Fig.(\ref{fig3}b) will be realized by a monotonically decreasing $\lambda$ intersecting the $h_{\text{max}}$-branch of the CC twice, as is the case for the IDM model.

In conclusion, we propose that microscopic traffic models should start with identifying the proper optimal velocity function, from which the family of ground states are defined. Simple traffic models are constructed as special cases of the expansion around those ground states, with coefficients of expansion in general dependent on the vehicle density. In particular the ``synchronized phase" can be formally understood as the result of the models where the leading order expansion about the optimal velocity vanishes within some  density range. The theoretical framework also proposes standard ways of model construction, and validation by the empirical data. The understanding of various possible solutions to a more general class of OV models could also be useful for designing the algorithms of autonomous driverless vehicles for large scale highway transportation.

\begin{acknowledgements}
We thank Prof. Dirk Helbing for useful discussions. This research was partially supported by Singapore A$^{\star}$STAR SERC ``Complex Systems" Research Programme grant 1224504056.  
\end{acknowledgements}


\begin{thebibliography}{99}

\bibitem{helbing_crowd}
D. Helbing, L. Buzna, A. Johansson and T. Werner, Transport Sci. {\bf 39}, 1 (2005).

\bibitem{helbing_nature}
D. Helbing, I. Farkas and T. Vicsek, Nature {\bf 407}, 487 (2000).

\bibitem{dogbe}
N. Bellomo and C. Dogbe, Soc. Ind. App. Math. {\bf 53}, 409 (2011).

\bibitem{as}
D. Chowdhury, L. Santen and A. Schadschneider, Phys. Rep. {\bf 329}, 199 (2000).

\bibitem{helbing_control}
D. Helbing and A. Mazloumian, Eur. Phys. J. B. {\bf 70}, 257 (2009).

\bibitem{helbing}
D. Helbing, Rev. Mod. Phys. {\bf 73}, 1067 (2001).

\bibitem{kernerbook}
B.S. Kerner, Introduction to Modern Traffic Flow Theory and Control: The Long Road to Three-phase Traffic Theory, Springer-Verlag Berlin Heidelberg 2009, and the references therein. 

 \bibitem{kernerexp1}
B.S. Kerner, Phys. Rev. E. {\bf 65}, 046138 (2002). 
\bibitem{kernerexp2}
B.S. Kerner, Phys. Rev. Lett. {\bf 81}, 3797 (1998).

\bibitem{kernercrit}
B.S. Kerner, Physica A, {\bf 392}, 5261(2013). 

\bibitem{treiberbook}
M. Treiber and A. Kesting, Traffic Flow Dynamics, Springer-Verlag Berlin Heidelberg 2013, and the references therein. 
\bibitem{helbingcrit}
M. Treiber, A. Kesting and D. Helbing, Transport. Res. B. {\bf 44}, 983 (2010). D. Helbing, M. Treiber, A. Kesting and M. Schonhof, Eur. Phys. J. B. {\bf 69}, 583 (2009). One should also note that empirically the spatiotemporal patterns of the synchronized phase fluctuates over short time and distance. 

\bibitem{helbing_idm}
M. Treiber, A. Hennecke and D. Helbing, Phys. Rev. E. {\bf 62} 1805 (2000).

\bibitem{kerner_mod1}
B.S. Kerner and S.L. Klenov, J. Phys. A: Math. Gen. {\bf 35} L31-L43 (2002).
B.S. Kerner and S.L. Klenov, J. Phys. A: Math. Gen. {\bf 39} 1775 (2006).

\bibitem{kerner_mod2}
B.S. Kerner and S.L. Klenov, Phys. Rev. E. {\bf 68}, 036130 (2003).

\bibitem{yangbo}
Bo Yang, Xihua Xu, John Z.F. Pang and Christopher Monterola, arXiv. 1504.01256. While the general framework does not favor one traffic model over the other, the verification of the master model by the experimental data allows us to justify the approximations employed by the model.

\bibitem{bando}
M. Bando, K. Hasebe, A. Nakayama, A. Shibata, and Y. Sugiyama, Phys. Rev. E. {\bf 51}, 1035 (1995).

\bibitem{JiangR_PRE01}
R. Jiang, Q. Wu and Z. Zhu, Phys. Rev. E. {\bf 64}, 017101 (2001).

\bibitem{GLW_PhyA08}
H. Gong, H. Liu and B. Wang, Phys. A. {\bf 387}, 2595 (2008).

\bibitem{shamoto}
D. Shamoto, A. Tomoeda, R. Nishi and K. Nishinari, Phys. Rev. E. {\bf 83}, 046105 (2011).

\bibitem{fn2}
If $\lambda\left(h_0\right)>2\text{sech}(h_0)$ but below the coexistence curve, the $h_n=\text{const}$ solution can be absolutely stable or unstable depending on the detailed form of $\lambda$.  It is stable if the oscillations of partially developed cluster solution are small, since a large enough perturbation is needed to destabilize the $h_n=\text{const}$ solution\cite{kerner_phase}. 

\bibitem{kerner_phase}
B.S. Kerner and P. Konhauser, Phys. Rev. E. {\bf 50}, 54 (1994).

\bibitem{yu_phase}
W. Shi, N-G Chen and Y. Xue, Commun. Theor. Phys. {\bf 48}, 1088 (2007).

\bibitem{naka_phase}
H. Hayakawa and K. Nakanishi, Phys. Rev. E. {\bf 57}, 3839 (1998).

\bibitem{nagatani_phase}
T. Nagatani, Phys. Rev. E. {\bf 61}, 3564 (2000).

\bibitem{xue_phase}
H.X. Ge, S.Q. Dai, L.Y. Dong and Y. Xue, Phys. Rev. E. {\bf 70}, 066134 (2004).

\bibitem{fn}
Both of them depend only on $\lambda$ and \emph{independent} of the vehicle density. In this special case the two branches are symmetric about the y-axis. Also see arXiv. 1407.3177.



\end{thebibliography}
\end{document}